\documentclass[aps,twocolumn,showpacs,preprintnumbers,amsmath,amssymb,floats,citeautoscript,nobalancelastpage]{revtex4-2}
\usepackage{graphicx}
\usepackage{dcolumn}
\usepackage{bm}
\usepackage{txfonts}
\usepackage{multirow}
\usepackage{color}

\begin{document}
 
\title{Prominent electron-hole asymmetry in thermoelectric transport of LaCoO$_3$}

\author{Ryuji~Okazaki$^1$}
\email{okazaki@rs.tus.ac.jp}
\author{Keisuke~Tomiyasu$^{2,3}$}

\affiliation{$^1$Department of Physics, Faculty of Science and Technology, Tokyo University of Science, Noda 278-8510, Japan}
\affiliation{$^2$Department of Physics, Tohoku University, Sendai 980-8578, Japan}
\affiliation{$^3$Nissan ARC Ltd., Natsushima-cho 1, Yokosuka 237-0061, Japan}

\begin{abstract}
We have measured the electrical resistivity and the thermopower of the electron-doped perovskite cobaltites LaCo$_{1-y}$Te$_y$O$_3$.
In contrast to the hole-doped systems such as metallic ferromagnets La$_{1-x}M_x$CoO$_3$ ($M$ = Ca, Sr, Ba),
the electron-doped samples show an insulating behavior even in a heavily doped range
due to a spin-state blockade mechanism that an electron hopping from high-spin Co$^{2+}$ to low-spin Co$^{3+}$ site
is energetically suppressed.
We find that, despite the electron doping, the thermopower
shows relatively large positive values above $y=0.05$,
strikingly distinct from the hole-doped case where it comes close to zero with doping.
This prominent electron-hole asymmetry seen in the thermopower originates from a bipolar conduction
which consists of a slight amount of mobile holes and the main immobile electrons,
demonstrating an impact of spin-state blockade on thermoelectric transport.

\end{abstract}

\maketitle

\section{introduction}

An electron-hole symmetry, 
which treats
an electron hole as a positively-charged quasiparticle
on an essential equality with the electron itself,
is a fundamental concept to construct the theory of solids \cite{AM},
as is widely seen in various systems of contemporary condensed matter physics \cite{Sato2001,Herrero2004,Yin2008}.
Such an equivalence between electrons and holes
approximately holds in metals 
owing to a small energy dependence of the density of states near the Fermi level.
On the other hand, the electron-hole symmetry is generally
absent in semiconductors,
mainly due to the difference in
the \textit{orbital} characters of conduction and valence bands.

An exotic origin to induce the electron-hole asymmetry comes from 
the \textit{spin} sector of the internal degrees of freedom of electrons,
and 
a peculiar class of Co oxides offers an intriguing playground to examine such a 
spin-driven asymmetry \cite{Maignan2004}.
In CoO$_6$ octahedron,
fivefold degenerate Co $3d$ orbitals split into twofold $e_g$ and threefold $t_{2g}$ levels,
and the spin state can be sensitively varied due to 
delicate energy balance among 
the crystal-field splitting and the Hund coupling;
for instance, in a Co$^{3+}$ ion,
$t_{2g}$ orbitals are fully occupied by
six $d$ electrons to form low-spin (LS) configuration ($e_g^0t_{2g}^6$, $S=0$)
when the crystal-field splitting is larger than the Hund coupling. 
For the opposite situation, 
Co$^{3+}$ ions take high-spin (HS) state ($e_g^2t_{2g}^4$, $S = 2$).
Most interestingly, such spin configurations significantly affect 
 hopping probability of correlated carriers.
As schematically shown in Fig. 1(a), 
LS Co$^{3+}$ ion can not accept electron hopping from
HS Co$^{2+}$ ion
owing to highly-unstable resulting LS Co$^{2+}$ state,
which can be referred as spin-state blockade.
On the other hand, hole hopping from LS Co$^{4+}$ ion is acceptable as drawn in Fig. 1(b),
leading to an electron-hole asymmetry in the conduction process.

\begin{figure}[b]
\includegraphics[width=1\linewidth]{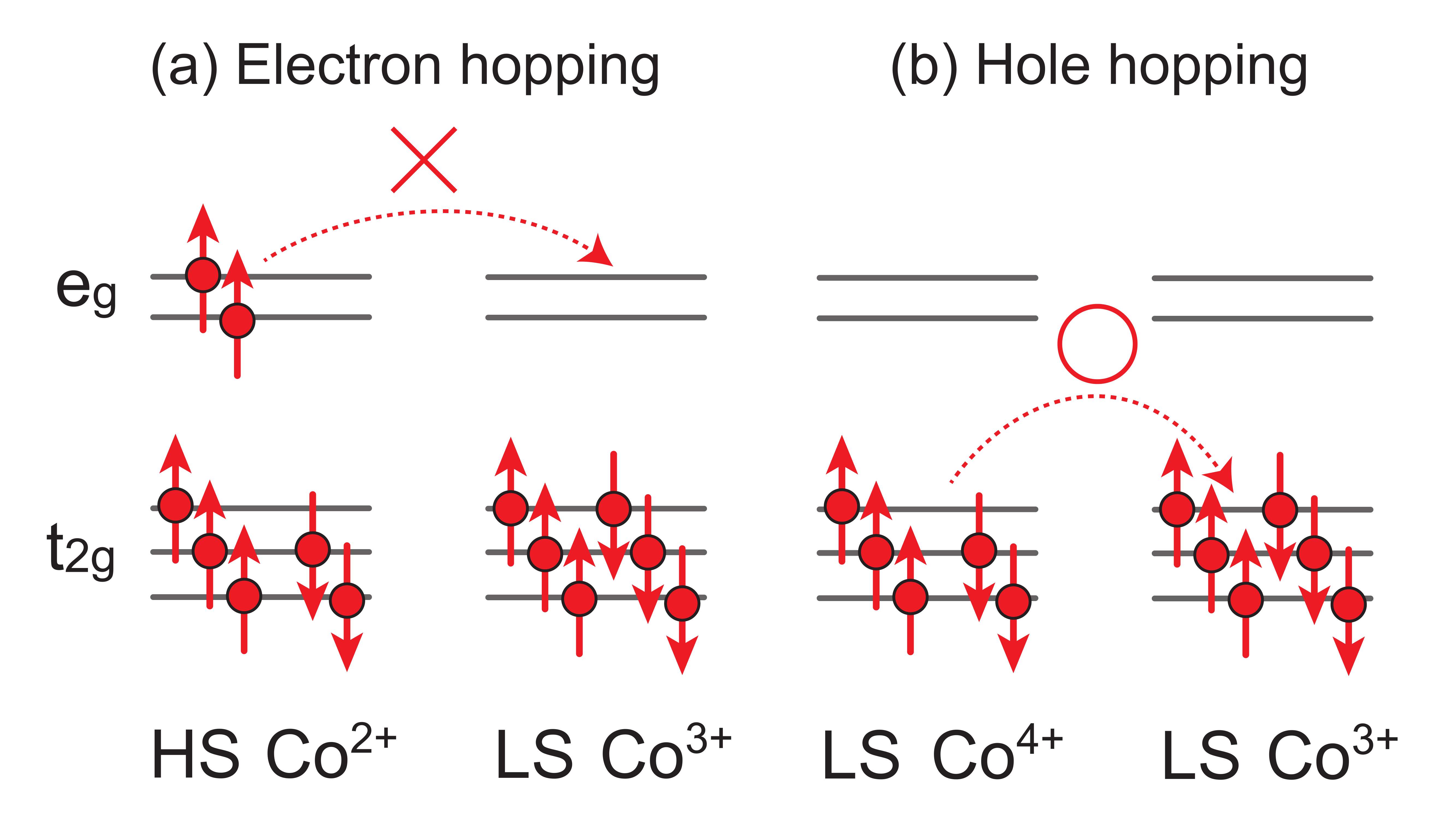}
\caption{
Schematic figures of (a) an electron hopping from high-spin (HS) Co$^{2+}$ to low-spin (LS) Co$^{3+}$ site and
(b) a hole hopping from LS Co$^{4+}$ to LS Co$^{3+}$ site.
Electron hopping into LS Co$^{3+}$ site is not allowed because the resultant LS Co$^{2+}$ is 
highly unstable.
}
\end{figure}

Materials realization of such spin-state-induced asymmetry has been discussed in 
several Co oxides \cite{Maignan2004,Taskin2005PRL,Taskin2005PRB,Taskin2006,Chang2009,Singh2019}.
Among them, 
a perovskite cobaltite LaCoO$_3$ may serve as an ideal case 
for investigation of the asymmetric transport in the background of LS Co$^{3+}$,
as is firstly claimed by Maignan \textit{et al.} \cite{Maignan2004}.
Indeed, 
crystal structure of LaCoO$_3$ consists of a simple three-dimensional network of 
corner-shared CoO$_6$ octahedra,
in contrast to rather complicated structures of the other cobaltites reported in earlier studies
such as oxygen-deficient perovskites \cite{Maignan2004,Taskin2005PRL,Taskin2005PRB,Taskin2006}
and charge-ordered double-perovskite oxides \cite{Chang2009}.
On the doping effect, 
although hole doping has been intensively investigated 
in La$_{1-x}A_x$CoO$_3$ ($A=$ Ca, Sr, Ba) \cite{Jonker1953,Itoh1994,Ganguly1994,Sehlin1995,Yamaguchi1996LCO,Wu2003,Kriener2004,Berggold2005,Iwasaki2008,Wang2010,Long2011}, 
 electron doping effect, 
which could be achieved by substituting tetravalent ions  ($A=$ Ce, Th, Te),
is less studied mainly due to difficulties in the sample preparations;
single-phase bulk of Ce-substituted compounds is difficult to obtain \cite{Tabata1989,Maignan2004Ce,Fuchs2005} 
and 
Th is radioactive \cite{Tabata1987}.
The structural and magnetic properties of La$_{1-x}$Te$_x$CoO$_3$ are available \cite{Zheng2008,Thakur2018},
yet microscopic valence state of Co ions is unclear.
The electron doping effect is mostly demonstrated in LaCo$_{1-y}$Ti$_y$O$_3$ \cite{Bahadur1983,Nakatsugawa1999,jirak2008,Serrano2008,Hejt2008},
and a detailed transport measurement has been performed 
while the systematic results are limited at high-temperature regime with a small population of the LS Co$^{3+}$ ions \cite{Robert2010}, 
which are essential for the spin-state blockade.
Thus, 
the underlying transport mechanism that how does the spin-state blockade work in the electron-doped LaCoO$_3$ is still missing.

To examine the suggested electron-hole asymmetry in LaCoO$_3$,
we here report a systematic evolution of the thermopower in electron-doped LaCo$_{1-y}$Te$_y$O$_3$.
In this compound, 
 valence state of Te ions is confirmed as non-magnetic Te$^{6+}$ 
by microscopic experiments of x-ray absorption spectra (XAS) and electron spin resonance \cite{Tomiyasu2016}
as well as  macroscopic magnetization measurements \cite{Tomiyasu2017},
providing a suitable platform to study asymmetric electron transport.
Also, the thermopower is a sensitive probe to detect the electron-hole asymmetry \cite{Behnia}.
We find that
 electron-doped samples show an insulating resistivity even in a heavily doped range due to a spin-state blockade mechanism for electrons.
Remarkably, the thermopower of LaCo$_{1-y}$Te$_y$O$_3$
shows relatively large positive values above $y=0.05$ in spite of electron doping.
This is in sharp contrast to hole-doped case where it comes close to zero in a usual manner for carrier doping.
We suggest a bipolar conduction
composed of mobile holes and immobile electrons,
giving a clear picture for observed electron-hole asymmetry in the thermopower.

\section{experiments}

Polycrystalline samples of LaCo$_{1-y}$Te$_y$O$_3$ ($y=0$, 0.01, 0.03, 0.05, 0.10, 0.15) were 
synthesized by a conventional solid-state reaction \cite{Tomiyasu2016,Tomiyasu2017}. 
The starting materials, La$_2$O$_3$ (99.99\% purity),
Co$_3$O$_4$ (99.9\%), and TeO$_2$ (99.9\%) powders, were 
mixed in the stoichiometric ratio, ground throughly, pelletized, and heated on ZrO$_2$ plates at 1473 K in air for 12 h twice (24 h in total).
No impurity phases were detected in the x-ray diffraction measurements. 

The resistivity was measured using a standard four-probe method.
The excitation current of $I=10$~$\mu$A was provided by a Keithley 6221 current source and 
the voltage was measured with a synchronized Keithley 2182A nanovoltmeter 
using a built-in Delta mode to cancel thermoelectric voltage.
The thermopower was measured using a steady-state technique
with a typical temperature gradient of 0.5 K/mm made by a resistive heater.
The thermoelectric voltage of the sample was measured with Keithley 2182A nanovoltmeter.
The temperature gradient was measured with
a differential thermocouple made of copper and constantan in a liquid He cryostat below 300~K and
with two platinum temperature sensors in a high-temperature equipment from 300~K to 700~K \cite{Ikeda2016,Sakabayashi2021}.
In both measuring apparatuses,
the thermoelectric voltage from the wire leads was subtracted.
A part of the data is lacking near room temperature, which is just the gap between the low- and the high-temperature apparatuses.

\section{results and discussion}

\begin{figure}[t]
\includegraphics[width=1\linewidth]{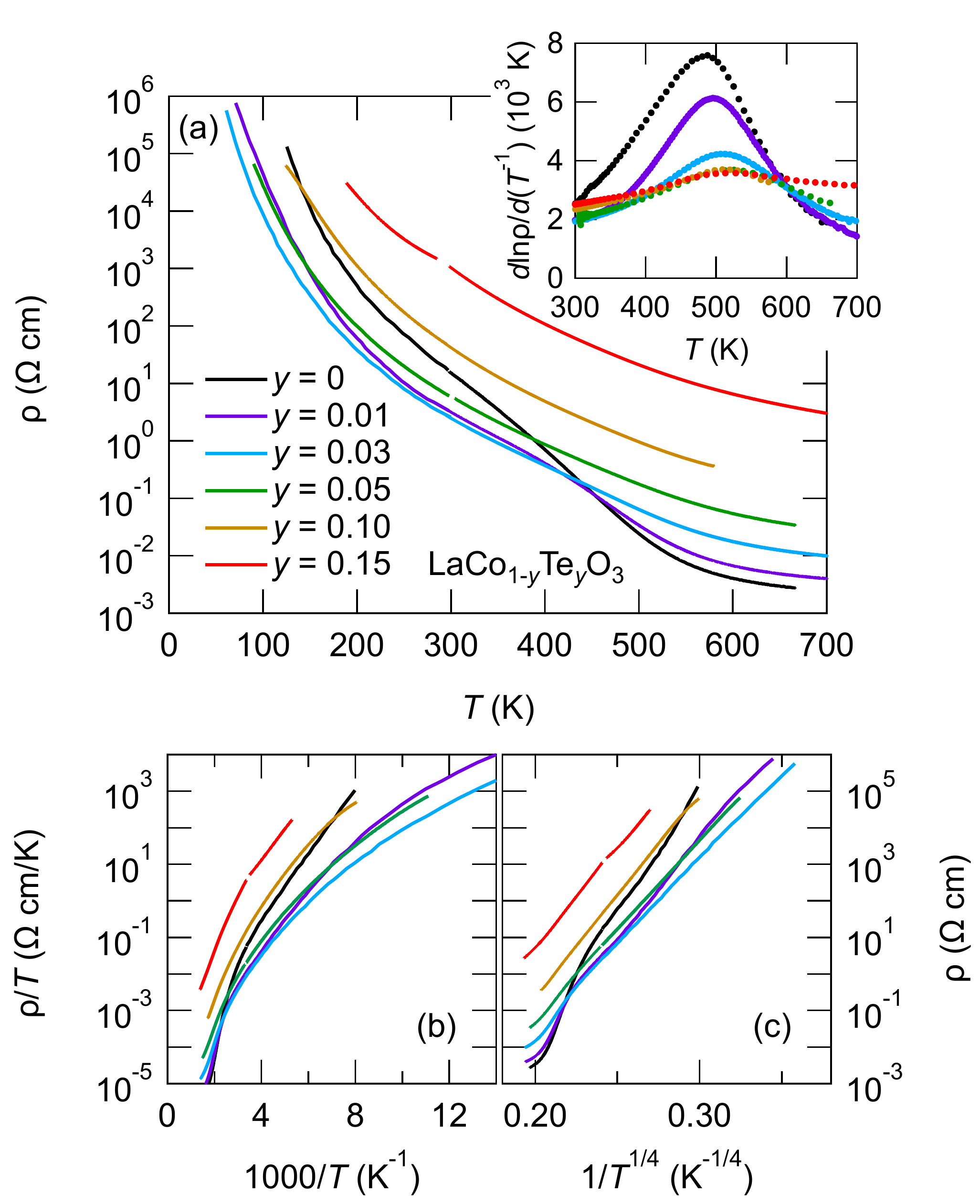}
\caption{
(a) Temperature variations of the resistivity $\rho$ of polycrystalline LaCo$_{1-y}$Te$_y$O$_3$ ($y=0$, 0.01, 0.03, 0.05, 0.10, 0.15).
The inset depicts the temperature dependence of $d\ln\rho/d(T^{-1})$ above room temperature.
(b) $T^{-1}$ dependence of $\rho/T$ and (c) $T^{-1/4}$ dependence of $\rho$.
}
\end{figure}

Figure 2(a) represents temperature variations of the electrical resistivity in
polycrystalline LaCo$_{1-y}$Te$_y$O$_3$.
In the earlier report \cite{Tomiyasu2016},
the resistivity was measured only below room temperature and the Te content $y$ was limited to
$0\leq y\leq0.05$.
Here we find that 
temperature dependence of the resistivity
is insulating for all the Te-substituted samples up to $y=0.15$
in spite of electron doping by Te$^{6+}$ substitutions.
It should noted that present results are clearly distinct from hole doping effect 
in La$_{1-x}A_x$CoO$_3$ ($A=$ Ca, Sr, Ba),
in which  metallic temperature dependence of the resistivity is observed for heavily-doped samples \cite{Kriener2004}.
Therefore, the insulating behavior in present electron-doped LaCoO$_3$ indicates that the electron mobility is significantly low,
as is expected in the spin-blockade scheme shown in Fig. 1 \cite{Maignan2004}.
Together with the thermopower results, detailed comparison between  electron- and  hole-doping effects 
on the transport properties will be discussed later.

In LaCoO$_3$, 
the resistivity significantly decreases with heating across a crossover temperature $T_{\rm MI} \approx 530$~K
from low-temperature insulating to high-temperature metallic states,
at which the specific heat exhibits a broad peak \cite{Stolen1997}.
This crossover is seen in temperature variations of 
$d\ln\rho/d(T^{-1})$,
which would be equivalent to an activation energy $\Delta/k_{\rm B}$ 
for the activation-type resistivity of $\rho(T) = \rho_0\exp(\Delta/k_{\rm B}T)$
if the activation energy is temperature-independent 
($\rho_0$ and $k_{\rm B}$ being a high-temperature extrapolation and the Boltzmann constant, respectively) \cite{Yamaguchi1996}.
The inset of Fig. 2(a) represents the temperature variations of 
$d\ln\rho/d(T^{-1})$ in LaCo$_{1-y}$Te$_y$O$_3$.
A pronounced peak corresponding to the crossover is clearly seen in the parent compound, 
and although the peak is smeared, 
it seems to shift slightly to higher temperature with Te substitutions.
This tendency is similar to that in electron-doped LaCo$_{1-y}$Ti$_y$O$_3$ \cite{jirak2008},
and may be consistent with a picture that
 doped electrons are immobile due to the blockade to enhance insulating nature.

Below room temperature, the resistivity of LaCoO$_3$ has been analyzed 
in terms of a small polaron hopping model \cite{Iguchi1996}, 
in which the resistivity is given as $\rho(T) \propto T\exp[(W_{\rm H}+E_{\rm g}/2)/k_{\rm B}T]$,
where $W_{\rm H}$ and $E_{\rm g}$ are  hopping energy of a polaron and the band gap, respectively.
In Fig. 2(b), we show $T^{-1}$ dependence of $\rho/T$ for LaCo$_{1-y}$Te$_y$O$_3$.
Although $\rho/T$ in pure LaCoO$_3$ behaves a linear variation in the log plot as expected in the small polaron model,
$\rho/T$ data in Te-substituted LaCoO$_3$ are convex upward,
indicating a considerable disorder effect.
Indeed, in earlier study \cite{Tomiyasu2016}, 
$\rho(T)$ in LaCo$_{1-y}$Te$_y$O$_3$ ($0<y\leq 0.05$) is fitted by a power-law relation of 
$\rho(T)\propto 1/T^{\nu}$ ($\nu=8$ to 10), 
which is expected in a situation that carriers hop over barriers of variable heights \cite{Marshall2007},
in a similar manner to the case of LaCo$_{1-y}$Ti$_y$O$_3$ \cite{jirak2008}.
On the other hand, the resistivity in such a disordered system is often analyzed by adopting a 
variable-range-hopping (VRH) formula of $\rho(T)\propto\exp(1/T)^{1/(1+d)}$,
where $d$ denotes the dimensionality \cite{Mott,Okuda2005}.
Figure 2(c) represents $T^{-1/4}$ dependence of $\rho$ for LaCo$_{1-y}$Te$_y$O$_3$,
in which one can see linear variations for Te-substituted samples, implying a 
validity of VRH conduction in three-dimensional systems.
In general, however, it is difficult task to uniquely determine the conduction mechanism from  
the resistivity; 
both power-law and VRH formulas seem to be applicable for present Te-substituted case,
and the disorder effect due to Te substitutions is essential for both conduction processes.

\begin{figure}[t]
\includegraphics[width=1\linewidth]{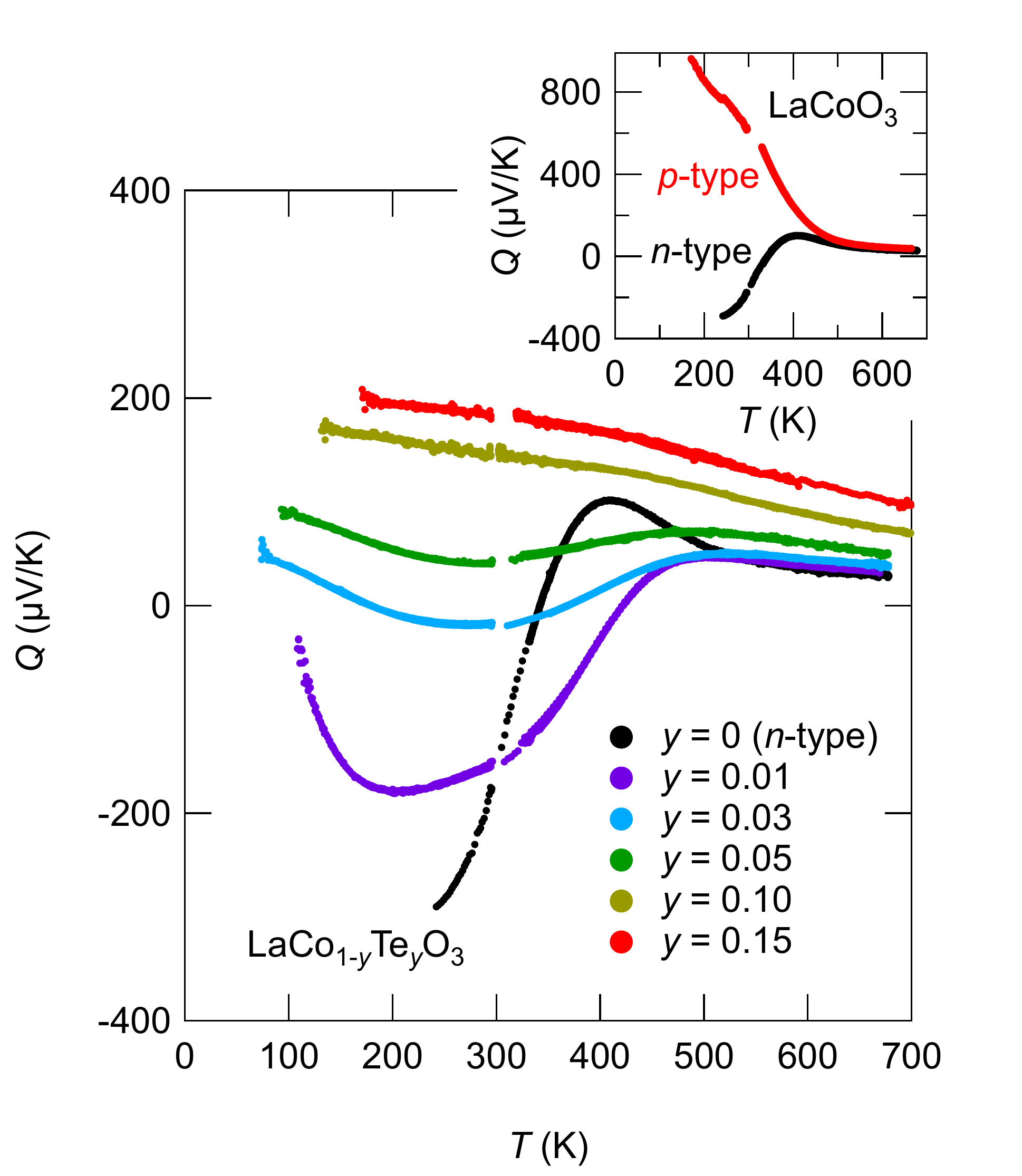}
\caption{
Temperature variations of the thermopower $Q$ in polycrystalline LaCo$_{1-y}$Te$_y$O$_3$.
The inset shows the sample dependence of the thermopower in LaCoO$_3$.
}
\end{figure}

Figure 3 displays the temperature variations of the thermopower in LaCo$_{1-y}$Te$_y$O$_3$
and the inset shows the sample dependence for the parent compound LaCoO$_3$.
In low-temperature insulating state of LaCoO$_3$, 
the sign of the thermopower has been reported to 
be either positive or negative \cite{Heikes1964,Sehlin1995},
which originates from a tiny amount of impurity.
This uncontrolled sample dependence is also seen in present study as shown in the inset.
Here, the temperature dependence of the thermopower and the electrical resistivity in LaCoO$_3$
is analyzed with a semiempirical bipolar model \cite{Sehlin1995}.
This model takes account of 
thermally-excited $p$- and $n$-type small polarons, 
which respectively correspond to Co$^{4+}$ and Co$^{2+}$,
with an activation energy
incorporating the Coulomb screening interaction.
Then, the conductivity $\sigma$ and the thermopower $Q$ are obtained using a 
bipolar model as 
$\sigma = \sigma_{p}+\sigma_{n}$ and 
$Q  = (\sigma_{p}Q_p+\sigma_{n}Q_n)/\sigma$,
where $\sigma_{i}$ ($Q_i$) $(i=p,n)$ is the conductivity (the thermopower) of $i$-type carrier,
by adopting a mobility ratio among the $p$- and $n$-type polarons.
Note that, 
although the thermopower $Q_i$ was calculated 
using the Heikes formula \cite{Chaikin1976},
which is particularly essential for Co oxides \cite{Terasaki1997,Koshibae2000,Hebert2021},
spin and orbital degeneracies were not involved in Ref. \onlinecite{Sehlin1995}.
Even by employing degeneracies, 
however, 
the observed thermopower may be reproduced 
by slight modification in the mobility ratio,
because it is given by an exponential form \cite{Sehlin1995} 
while 
the degeneracy is taken in the Heikes formula logarithmically.

We then focus on electron-doping effect on the thermopower.
As seen in Fig. 3, the thermopower dramatically changes with Te substitution,
and most interestingly,
the sign becomes to be positive for $y \geq 0.05$ 
in spite of  electron doping.
The present result is then distinct from
hole-doping effect observed in La$_{1-x}A_x$CoO$_3$ ($A=$ Ca, Sr, Ba).
To compare these doping effects,
we plot the resistivity and the thermopower measured at a constant temperature of $T=200$~K
as a function of the Co valence $\nu$ in Figs. 4(a) and 4(b), respectively.
Here we argue the 200-K data at which the considerable amount of LS Co$^{3+}$ ions exists to discuss the spin-state blockade as is also mentioned later.
Note that $\nu$ in LaCo$_{1-y}$Te$_y$O$_3$ is estimated as 
$\nu = (3-6y)/(1-y)$,
since valence state of Te$^{6+}$ is confirmed by XAS \cite{Tomiyasu2016}.
For hole-doping regime, 
data of La$_{1-x}$Sr$_x$CoO$_3$ are taken from Refs. \onlinecite{Berggold2005} and \onlinecite{Iwasaki2008}
and $\nu$ is estimated as $\nu = 3 + x$.

\begin{figure}[t]
\includegraphics[width=1\linewidth]{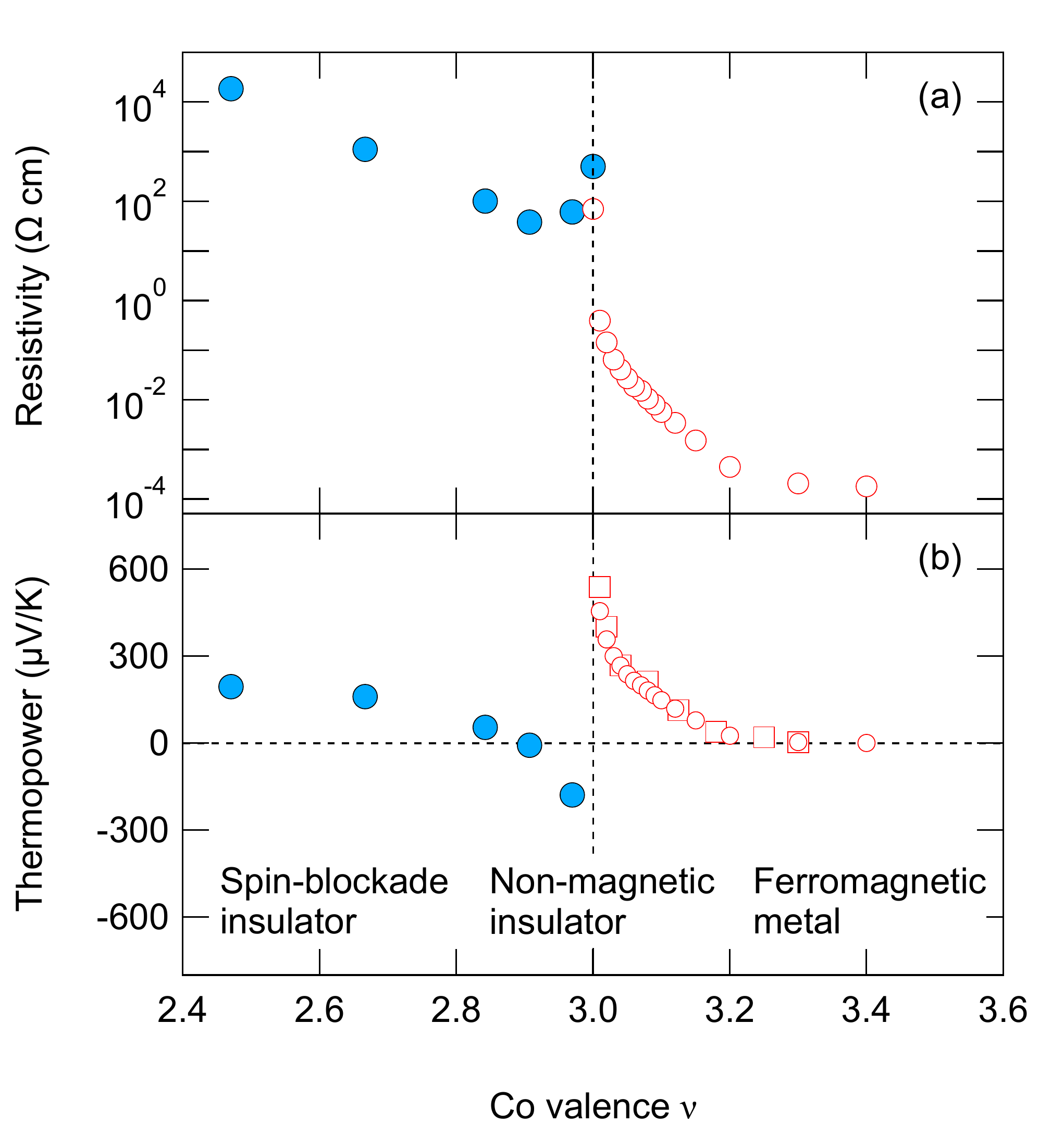}
\caption{
(a) The resistivity and (b) the thermopower measured at $T=200$~K as a function of Co valence $\nu$.
The present data of LaCo$_{1-y}$Te$_y$O$_3$ are plotted in the electron-doped regime as the filled circles.
In the hole-doped regime, 
the resistivity and the thermopower data of La$_{1-x}$Sr$_x$CoO$_3$ taken from Ref. \onlinecite{Iwasaki2008} are shown by the open circles.
The thermopower data depicted by the open squares are taken from Ref. \onlinecite{Berggold2005}.
}
\end{figure}

Now we discuss a prominent asymmetry between  electron- and hole-doping effects in LaCoO$_3$
as seen in Figs. 4(a) and 4(b).
In contrast to a conventional carrier doping effect in  hole-doped system, 
the resistivity initially decreases with Te substitutions ($y\leq 0.03$), but 
then increases above $y=0.05$,
indicating that 
 electrons are doped but the electron mobility is too low due to the aforementioned spin-state blockade.
We emphasize that 
the affinity of the HS Co ions favors the LS Co$^{3+}$ nearest neighbor surrounding \cite{Goodenough1,Goodenough2},
corroborating the spin-state blockade picture for the doped electrons [Fig. 1(a)].
Note that spin-state nature of LaCoO$_3$ in this temperature range near 200~K is a controversial issue;
the excited spin state is either intermediate spin (IS) \cite{Korotin1996,Mizokawa1996,Saitoh1997,Yamaguchi1997,Ishikawa2004} or 
HS state \cite{Noguchi2002,Ropka2003,Haverkort2006,Podlesnyak2006,Knizek2009,Asai2012,Tomiyasu2017rixs}.
It has been suggested that 
a collective heptamer picture involving IS and HS dual nature, which may also give a reasonable explanation to the polaron conduction, is essential \cite{Tomiyasuarxiv}.
On the other hand, even at 200~K, 
considerable amount of LS states, which are essential for the spin-state blockade, 
is remaining \cite{Tomiyasu2017rixs}, and this situation is the same for both doping cases.
It should also be noted that, although impurity effect is certainly seen in the resistivity,
concentration of scattering centers in Te-substituted compounds are much less than that
in the hole-doped system, 
because substituted Te ion is hexavalent to act as an effective electron donor.

The spin-state blockade for electrons may give a comprehensible explanation 
for observed sign change in the thermopower above $y=0.05$.
Based on the bipolar conduction in the parent compound \cite{Sehlin1995},
the thermopower is weighted by the conductivity of each carriers as $Q  = (\sigma_{p}Q_p+\sigma_{n}Q_n)/\sigma$.
In present electron-doped case, the electron mobility is significantly low 
and then 
the Peltier conductivity of holes may become relatively larger than that of electrons ($|\sigma_{n}Q_n| \ll |\sigma_{p}Q_p|$),
resulting in a positive thermopower possibly due to remnant minority holes, while the underlying nature of the minority holes is unclear at present.

We also discuss an entropy back-flow mechanism \cite{Kobayashi_eb} for the observed positive thermopower in the electron-doped region.
In a simple non-degenerate case,
an entropy of $k_{\rm B}\ln2$ (only spin degeneracy) is carried by a charge \cite{Behnia}.
On the other hand,
if the degeneracy of the destination site ($g_{\rm d}$) is larger than that of the initial site ($g_{\rm i}$)
in a local hopping picture,
the entropy flows back against the charge hopping [$k_{\rm B}\ln(g_{\rm i}/g_{\rm d})<0$].
As a result,
the sign of the thermopower becomes opposite to the sign of the charge of the carriers,
as reported in the Mn oxides \cite{Kobayashi_eb}.
In the electron-doped LaCoO$_3$,
we consider the electron hopping from HS Co$^{2+}$ to
the low-lying 
LS Co$^{3+}$ or
thermally-excited HS Co$^{3+}$ (or IS Co$^{3+}$) at $T=200$~K.
Here, the degeneracy of the initial site of HS Co$^{2+}$ is $g_{{\rm HS,Co}^{2+}}=12$.
Then, the degeneracy of the destination site is given as follows: 
For the LS Co$^{3+}$ case, 
the degeneracy of LS Co$^{3+}$ is unity ($g_{{\rm LS,Co}^{3+}}=1$),
and thereby the entropy flow is forward 
while the hopping probability is low due to the spin-state blockade.
On the other hand, the degeneracy of HS (IS) Co$^{3+}$
is $g_{{\rm HS,Co}^{3+}}=15$ ($g_{{\rm IS,Co}^{3+}}=18$), 
which is larger than the degeneracy of HS Co$^{2+}$ ($g_{{\rm HS,Co}^{2+}}=12$),
indicating an entropy back-flow. 
However, for HS (IS) Co$^{3+}$ case, 
the absolute value of back-flow thermopower is estimated as
$|(k_{\rm B}/e)\ln(g_{\rm i}/g_{\rm d})|\simeq19$~$\mu$V/K (35 $\mu$V/K) 
for $\nu=2.5$, 
at which Co$^{2+}$ and Co$^{3+}$ exist in equal amount.
This estimated value is much smaller than the observed thermopower of $Q\simeq200$~$\mu$V/K at $\nu=2.5$.
Hence, the back-flow mechanism is not solely responsible.
Here we keep a qualitative discussion since the temperature dependence of the thermopower is also complicated;
detailed consideration based on theoretical calculation,
along with further transport investigation such as the Hall effect, 
should be a future study.

Let us compare with earlier results on the Co oxides. 
While electron-doping effect on the thermopower has been tried to be investigated in La$_{1-x}$Ce$_x$CoO$_3$ \cite{Maignan2004Ce},
owing to the difficulty of the sample synthesis,
the number of data points is small to discuss the overall doping dependence.
Interestingly, the thermopower of LaCoO$_3$ is drastically changed by Ti$^{4+}$ substitution \cite{jirak2008},
while the doping dependence is limited.
The detailed carrier-doping effect has been explored in oxygen-deficient perovskites 
$R$BaCo$_2$O$_{5+\delta}$ ($R$ = Gd, Nd) \cite{Taskin2005PRL,Taskin2005PRB,Taskin2006}.
Indeed, the resistivity of the electron-doped $R$BaCo$_2$O$_{5+\delta}$ is too high,
similar to the present study,
indicating a universal electron blockade transport in the Co oxides consisting of LS Co$^{3+}$.
However, let us emphasize that 
the thermopower behavior in $R$BaCo$_2$O$_{5+\delta}$ is in total contrast to the present results;
the thermopower in $R$BaCo$_2$O$_{5+\delta}$ 
remains a negative value 
against the electron doping, 
and large absolute value is discussed in terms of the extended Heikes formula
rather than electron-hole asymmetry in the thermopower.
These results highlight that present Te-substituted LaCoO$_3$ is a minimal model to 
demonstrate electron-hole asymmetry caused by spin-state blockade for thermoelectric transport.

\section{Summary}

To summarize, systematic measurements of the resistivity and the thermopower on the electron-doped LaCo$_{1-y}$Te$_y$O$_3$ have been carried out.
In sharp contrast to hole-doped case,
the resistivity is significantly increased by electron doping. 
Furthermore,
the thermopower exhibits a prominent feature characterized by the sign change against electron doping.
These transport results underlie
electron-hole asymmetry due to spin-state blockade.

\section*{acknowledgements}
We thank Mr. Ken Yamashita for experimental assistance and
Mr. Shun-Ichi Koyama and Ms. Mika Sato for supporting the sample preparations.
This work was supported by the JSPS KAKENHI No. 18K03503.


\end{document}